\documentclass[12pt,a4paper]{article}
\usepackage[T1]{fontenc}
\usepackage[utf8]{inputenc}
\usepackage[affil-it]{authblk}
\usepackage{amsmath}
\usepackage{amssymb}
\usepackage{bm}
\usepackage{lmodern}
\usepackage{xcolor}
\usepackage{graphicx}
\usepackage{float}
\usepackage[font=small,labelfont=bf]{caption}
\usepackage{hyperref}
\hypersetup{
    colorlinks=true,
    linkcolor=blue,
    citecolor=blue,
    filecolor=blue,
    urlcolor=blue,
    pdftitle={Reservoir- and Measurement-free Microwave Initialization of Semiconductor Spin Qubits},
}
\usepackage{geometry}
\usepackage{lineno}
\usepackage{siunitx}

\newcommand{\ket}[1]{\left|#1\right\rangle}

\title{\large\textbf{Reservoir- and Measurement-free Microwave Initialization of Semiconductor Spin Qubits}}

\author[1,${\dagger}$]{Leon C. Camenzind}
\author[1]{Ik Kyeong Jin}
\author[1]{Akito Noiri}
\author[1]{Kenta Takeda}
\author[1]{Takashi Nakajima}
\author[2]{Takashi Kobayashi}
\author[1,2,${\dagger}$]{ Seigo Tarucha}

\affil[1]{Center for Emergent Matter Science, RIKEN, 2-1 Hirosawa, Wako-shi, 351-0198, Saitama, Japan}
\affil[2]{Center for Quantum Computing, RIKEN, 2-1 Hirosawa, Wako-shi, 351-0198, Saitama, Japan}
\affil[$\dagger$]{Correspondence should be addressed to L.C.C.\ (\texttt{leon.camenzind@riken.jp}) or S.T.\ (\texttt{tarucha@riken.jp}).}
\date{29 / 9 / 2026}

\begin{document}
\maketitle

\begin{abstract}
\small
Scalable quantum processors require repeated qubit initialization throughout large arrays. In semiconductor spin qubits, fast initialization commonly relies on local reservoir access or measurement-based feedback, requiring dedicated infrastructure that becomes increasingly difficult to distribute as processors scale. Here, we demonstrate reservoir- and measurement-free initialization of a silicon spin-qubit pair in an industrially fabricated Si/SiGe quantum-dot device using a fixed sequence of microwave and baseband pulses. Odd spin-parity states relax to the singlet charge state, whereas blocked even spin-parity states are microwave-driven through the triplet manifold and subsequently converted to the singlet by singlet-triplet mixing and charge hybridization. Repeated cycles produce the singlet-associated charge outcome with a median probability of 99.4\% across the sampled preparation states, while exchange spectroscopy independently verifies mapping to the target $\ket{\uparrow\downarrow}$ operational state. Microwave spectroscopy and time-domain measurements identify the dark-state-limited single-cycle transfer and the subsequent blockade-lifting dynamics that set the initialization time scale. The demonstrated pumping sequence uses approximately \SI{12}{\micro\second} of microwave bursts and mixing dwells, while we project sub-microsecond initialization under improved device conditions. These results establish fixed-sequence microwave initialization as a scalable control primitive for semiconductor spin-qubit processors, based on singlet-triplet physics that can be adapted to platforms with suitable Pauli-blockade transitions.
\end{abstract}

\newpage
\section{Introduction}

Silicon spin qubits combine small device footprints, long coherence times and compatibility with industrial semiconductor processing, making them a promising route toward large-scale quantum processors~\cite{loss_quantum_1998,Ladd2010-tf,zwerver_qubits_2022,camenzind_spin_2022,weinstein_universal_2023,Steinacker2024-xu,neyens_probing_2024}. Recent experiments have established high-fidelity single-qubit gates, two-qubit gates, multi-qubit operation and advanced readout in silicon devices~\cite{yoneda_quantum-dot_2018,noiri_fast_2022,xue_quantum_2022,mills_two-qubit_2022,philips_universal_2022,Takeda2024-fm,Wu2025-highfidelity}. As arrays grow, however, initialization becomes an architectural constraint: providing reservoir and measurement infrastructure at every qubit location becomes increasingly demanding as processor size grows, particularly for qubits deep within dense two-dimensional arrays. Scalable initialization should therefore remain possible in both dense and sparse architectures without requiring a local reservoir or charge sensor (Fig.~\ref{fig:fig1}a).

Intrinsic spin relaxation is generally too slow, typically on millisecond-to-second timescales, to serve as an in-sequence reset step for cycle times of order microseconds. Faster initialization commonly uses energy-selective tunnelling between a quantum dot and an electron reservoir. This principle underlies single-spin initialization in Elzerman-type readout and related initialization schemes~\cite{elzerman_single-shot_2004}, as well as two-electron spin-singlet loading in double quantum dots~\cite{Takeda2024-fm}. Reservoir-based initialization is effective and can be fast~\cite{Blumoff2022,Takeda2024-fm}, but requires reservoir access at each initialization site. In common accumulation-mode devices, the electron accumulation regions connecting ohmic contacts to the dots, together with their control gates, consume area and routing resources. Providing this access to interior qubits in dense two-dimensional arrays can therefore strongly constrain the device layout~\cite{Vandersypen_interfacing_2017,boter_spiderweb_2022,Li2025-yd}.

A second possible route is active measurement-based feedback~\cite{philips_universal_2022,kobayashi_feedback-based_2023} (Fig.~\ref{fig:fig1}a). In feedback-based reset, the parity is measured in Pauli-spin blockade readout, and a conditional spin flip is applied if the spin parity is even ($\ket{\downarrow\downarrow}$ or $\ket{\uparrow\uparrow}$), effectively lifting the blockade and initializing the system into the singlet state. More generally, measurement- and logic-assisted initialization protocols can prepare high-fidelity two-qubit states even when thermal polarization is weak~\cite{Huang2024}. Such approaches can provide high-fidelity initialization, but require fast measurement and real-time discrimination. Conventional charge sensors require additional gates and commonly reservoir connectivity, making their placement in the interior of dense two-dimensional arrays similarly restrictive.

Shuttling and SWAP operations can transport qubits to shared initialization zones and thereby reduce the density of reservoir and sensor infrastructure, but initialization must still be provided at those zones, and accessing them adds routing, control and timing overhead~\cite{boter_spiderweb_2022,Li2025-yd}. A fixed-sequence local reset can therefore be used directly at an operation zone or provide added flexibility within a shuttling-based architecture.

This motivates fixed-sequence initialization through internal spin-charge conversion, without reservoir access or measurement feedback~\cite{Friesen2004}. Such control can be implemented locally, as demonstrated here, or potentially through globally applied microwave fields \cite{Vahapoglu2022} combined with local qubit selection.

Here we demonstrate microwave-based initialization of a spin pair in a Si/SiGe quantum-dot spin qubit device with micromagnets~\cite{yoneda_quantum-dot_2018}. The protocol resets the spin pair using its existing microwave and baseband controls, without electron tunnelling to a reservoir or measurement-conditioned feedback. The nearby charge sensor is used only to characterize and verify the protocol. The protocol exploits the spin-charge structure near the $(1,3)$--$(0,4)$ transition, combining selective microwave excitation of blocked spin population with Pauli-spin-blockade-mediated conversion into the singlet ground state. The relevant states are the four $(1,3)$ product states, $\ket{\downarrow\downarrow}$, $\ket{\uparrow\downarrow}$, $\ket{\downarrow\uparrow}$ and $\ket{\uparrow\uparrow}$, together with the $(0,4)$ singlet and triplet states. We refer to the parallel-spin states as even and the antiparallel-spin states as odd.

The initialization cycle uses the different roles of these states. Odd states relax into $\ket{(0,4)S}$ in the Pauli-spin-blockade regime, whereas blocked even states are first transferred by microwave excitation through the $(0,4)$ triplet manifold. The resulting $T_0$-like population is then converted to $\ket{(0,4)S}$ by singlet-triplet mixing and charge hybridization. Population reaching $\ket{(0,4)S}$ is no longer resonantly driven by the microwave pulse, so repeated cycles accumulate population in this reset state. From $\ket{(0,4)S}$, the spins are brought into $\ket{\uparrow\downarrow}$ via high-fidelity diabatic and adiabatic passages for subsequent qubit operations~\cite{Takeda2024-fm}. Because the method relies on singlet-triplet physics in Pauli spin blockade, the same initialization principle is applicable to other semiconductor spin-qubit platforms with analogous Pauli-blockade physics.

We first establish the relevant spin-charge spectrum and identify the microwave transitions with a calibrated eight-state Hamiltonian. We then isolate the coherent triplet-transfer step and show that a dark state limits the ideal single-cycle conversion. Finally, we demonstrate repeated initialization above 99\%, verify initialization into the operational $\ket{\uparrow\downarrow}$ state, and characterize the conversion dynamics using repeated-cycle models.

\section{Results}
\subsection{Concept and spin-parity readout}
\begin{figure}[H]
 \centering
 \captionsetup{font=footnotesize,skip=0pt,width=1\linewidth}
 \includegraphics[width=1\textwidth]{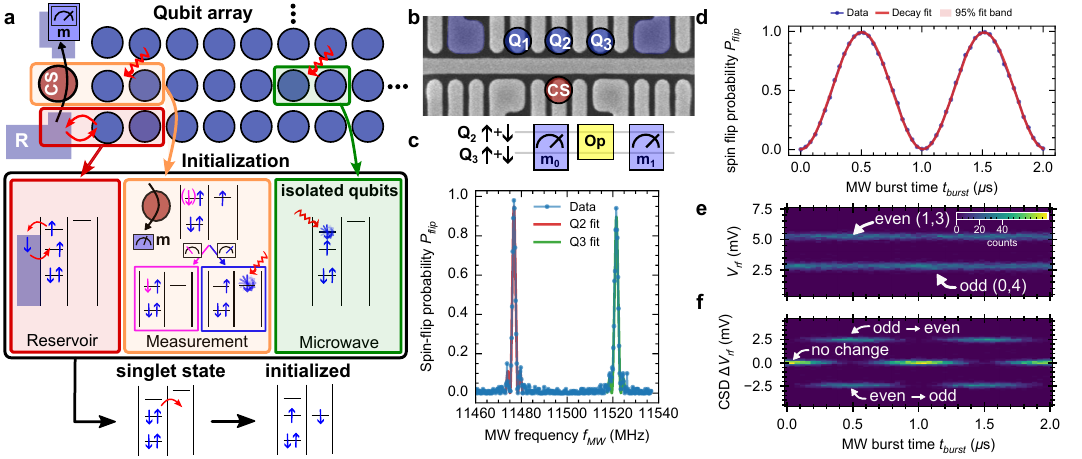}
 \caption{\textbf{Initialization concepts, device and parity readout.}
\textbf{a}, Schematic comparison of reservoir-, measurement- and microwave-based initialization. Reservoir-based initialization replaces an excited-spin electron with a ground-state electron through energy-selective tunnelling. In measurement-based initialization, an odd-parity outcome (pink) requires no spin flip, whereas an even-parity outcome (blue) triggers a conditional $\pi$ pulse. Microwave-based initialization operates without a local reservoir or charge sensor. All three routes prepare the singlet state, which is subsequently mapped to the operational initialized state. $R$ denotes a reservoir, $CS$ a charge sensor, and $m$ a single-shot measurement. Red arrows indicate electron tunnelling, while the red pulse symbol denotes qubit-control pulses.
 \textbf{b}, Si/SiGe three-qubit device with micromagnets (not shown), a charge sensor and the Q$_2$-Q$_3$ subsystem used here.
 \textbf{c}, EDSR spectroscopy of Q$_2$ and Q$_3$ at \SI{0.35}{T}, showing a frequency separation of \SI{44.8}{MHz} at the operation point.
\textbf{d}, High-visibility Rabi oscillation of Q$_2$, with a fitted initial contrast of approximately 99\%.
\textbf{e}, Histograms of single-shot charge-sensing events \(V_\mathrm{rf}\). Without initialization, flipping Q$_2$ in each cycle produces equal even-parity \((1,3)\) and odd-parity \((0,4)\) charge outcomes.
\textbf{f}, Charge-signal difference (CSD) computed from adjacent single-shot measurements in \textbf{e}, resolving odd-to-even and even-to-odd spin-flip events used to obtain the Rabi data in \textbf{d}.
}
\label{fig:fig1}
\end{figure}

Figure~\ref{fig:fig1} summarizes the initialization concepts and introduces the device and parity-readout used throughout the experiment. The measurements are performed in a three-qubit Si/SiGe device fabricated by Intel \cite{neyens_probing_2024, George2025-12qarrays}. We use Q$_2$ and Q$_3$ for the initialization protocol and detect the Q$_2$-Q$_3$ charge state with a nearby charge sensor (CS in Fig.~\ref{fig:fig1}b). Micromagnets provide both qubit addressability and electric-dipole spin resonance (EDSR) when microwave signals are applied to the horizontal splitting gate in the center~\cite{Tokura2006,Golovach_2006_EDSR}. The charge state is measured with radio-frequency reflectometry~\cite{Noiri2020_RF} using a thin-film NbTiN inductor chip.

Before applying the initialization protocol, we characterize coherent control and readout. At an external magnetic field of \SI{0.35}{T}, the micromagnet gradient produces a Zeeman-frequency difference $\Delta f_Z=\SI{44.8}{MHz}$ between Q$_2$ and Q$_3$ at the operation point (Fig.~\ref{fig:fig1}c). A microwave pulse applied between two parity measurements, $m_0$ and $m_1$, produces a detectable change in the charge state, allowing the spin-flip probability to be extracted without prior initialization for tuning up the qubits. The pulses for the readout are optimized such that the detected spin-flip visibility is approximately 99\% for both qubits. The high visibility Rabi-oscillations of Q$_2$ in Fig.~\ref{fig:fig1}d confirm both coherent microwave control and a well-tuned Pauli-spin-blockade readout, with only a small probability ($\ll1\%$) for unwanted spin-flips during the baseband passages through the anticrossings~\cite{Takeda2024-fm}. Because these measurements are performed without prior initialization, odd and even states are approximately equally populated after operation (Fig.~\ref{fig:fig1}e). The charge-signal difference (CSD), obtained from the difference between the sensor records in $m_0$ and $m_1$, separates odd-to-even and even-to-odd transitions and therefore provides the diagnostic used in the measurements below (Fig.~\ref{fig:fig1}f).

\subsection{Spin-charge spectrum of the initialization cycle}
\begin{figure}[H]
 \centering
 \captionsetup{font=footnotesize,skip=0pt,width=1\linewidth}
 \includegraphics[width=1\textwidth]{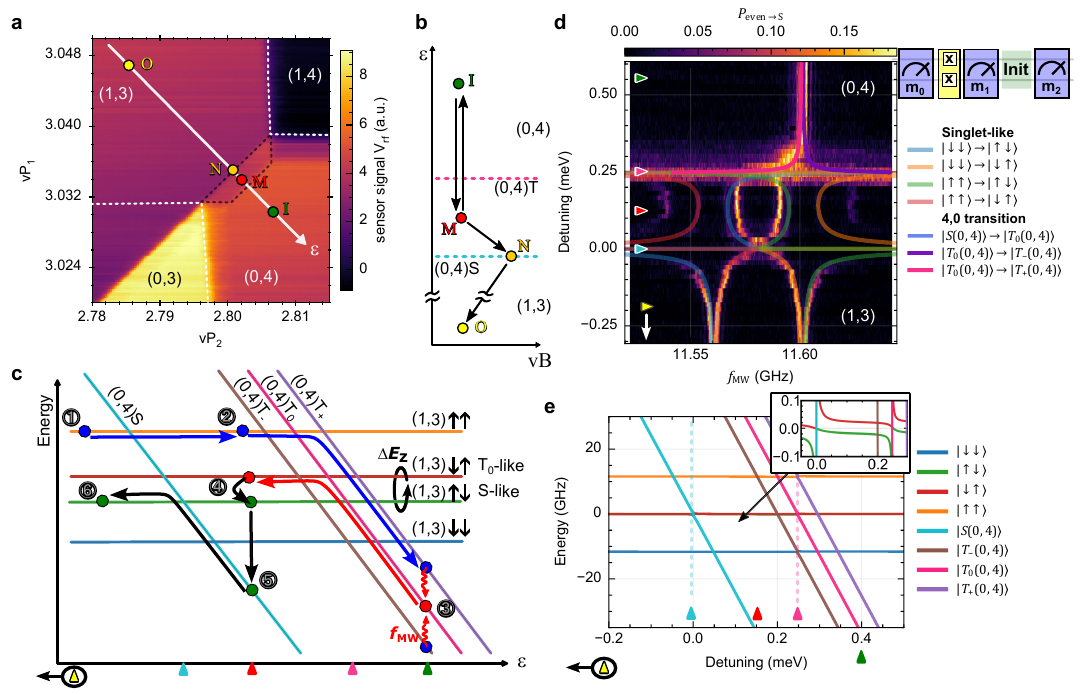}
 \caption{\textbf{Spin-charge spectrum underlying microwave initialization.}
 \textbf{a} Canonical charge stability diagram identifying the Pauli-spin-blockade region and the color-coded pulse points for operation (O), interdot transition (N), measurement (M) and initialization (I).
 \textbf{b} Pulse trajectory in detuning and interdot-barrier space. All gates involved are virtualized.
 \textbf{c}, Energy schematic of the initialization cycle. Odd states relax to the singlet charge state $\ket{(0,4)S}$ in the Pauli-spin-blockade regime. Even states enter the $(0,4)$ polarized triplet manifold, are driven by microwaves into a $T_0$-like branch, and are then converted to $\ket{(0,4)S}$ by singlet-triplet mixing and singlet charge hybridization. Diabatic (adiabatic) passage through the $\ket{(0,4)S}$-$\ket{\downarrow\downarrow}$ (-$\ket{\uparrow\downarrow}$) anticrossings initializes the qubits into  $\ket{\uparrow\downarrow}$.
 Valley-excited states, which duplicate the spin-charge spectrum at shifted detuning, are omitted from the schematics for clarity.
 \textbf{d} Conversion probability as a function of microwave frequency and initialization detuning around the PSB regime. Model transition lines from the eight-state Hamiltonian identify the resonances that contribute to initialization. Initialization is performed at $\epsilon_\mathrm{I}\sim\SI{1}{meV}$, outside the plotted range. The green triangle indicates the regime of the initialization point.
 \textbf{e} Calculated energy spectrum. The main qubit splittings are set by the external field and micromagnet magnetization, while exchange and charge hybridization generate the smaller splittings relevant to the resonance positions. The inset shows a zoom-in of the odd-parity states.}
 \label{fig:fig2}
\end{figure}

The initialization protocol is implemented at the $(1,3)$-$(0,4)$ transition, shown in Fig.~\ref{fig:fig2}. This higher-occupancy Pauli-blockade transition still uses the two spin-$1/2$ qubits Q$_2$ and Q$_3$, with the other two electrons forming a spin-paired core, but provides a larger usable readout window than the conventional $(1,1)$-$(0,2)$ transition~\cite{philips_universal_2022}. Because the protocol relies on the local singlet-triplet structure, it is also applicable to conventional $(1,1)$-$(0,2)$ Pauli-blockade transitions. The charge stability diagram in Fig.~\ref{fig:fig2}a shows the canonical Pauli-spin-blockade region. Although this measurement retains finite reservoir coupling, the initialization cycle itself does not exchange electrons with the reservoir. The same protocol can therefore be performed on a completely isolated double-quantum-dot system. The qubits are operated at point O, in the center of the $(1,3)$ charge state, where the dots are weakly coupled and the exchange interaction is well below \SI{100}{kHz}. In this regime, Q$_2$ and Q$_3$ are accurately described as separate spins.

Figure~\ref{fig:fig2}b shows the pulse trajectory, Fig.~\ref{fig:fig2}c illustrates the initialization cycle, and Fig.~\ref{fig:fig2}e gives the energy spectrum calculated using parameters matched to the measured resonances in Fig.~\ref{fig:fig2}d (Methods). For readout, the system is pulsed from O through N to M in the Pauli-blockade window. The odd-parity states couple to $\ket{(0,4)S}$ through their singlet components. The Zeeman-frequency difference $\Delta f_Z$ mixes the $(1,3)$ singlet and $T_0$ components, allowing both odd-parity states to convert to the singlet charge state $\ket{(0,4)S}$, while the parallel-spin states remain blocked in $(1,3)$.

For initialization, the system is pulsed deep ($\epsilon_\mathrm{I}\sim\SI{1}{meV}$) into the $(0,4)$ charge state along the virtual detuning axis, crossing the $(0,4)$ triplet states. The absence of residual \((1,3)\)-like charge occupation deep in \((0,4)\) indicates that the blocked parallel-spin states are adiabatically mapped onto the corresponding \((0,4)\) polarized triplet branches, as illustrated in Fig.~\ref{fig:fig2}c. Assuming the odd states have efficiently converted to $\ket{(0,4)S}$, the remaining non-singlet population occupies the polarized triplets $\ket{(0,4)T_{-}}$ or $\ket{(0,4)T_{+}}$. At the initialization point I, a resonant microwave pulse drives transitions from these polarized triplets toward $\ket{(0,4)T_{0}}$, see Fig.~\ref{fig:fig2}c. For each repetition, the sequence then returns from M to I and back to M. On return to M, $\ket{(0,4)T_{0}}$ connects to a $T_0$-like $(1,3)$ state, represented in the schematic by the $\ket{\downarrow\uparrow}$ branch. During a mixing time $t_\mathrm{mix}=\SI{1}{\micro\second}$ at M, this state mixes with $\ket{(1,3)S}$ and converts into $\ket{(0,4)S}$ through the blockade-lifting process described above. Repeating the cycle, therefore, increases the probability of reaching the singlet charge state.

Figure~\ref{fig:fig2}d shows the even-to-singlet conversion probability $P_{\mathrm{even}\rightarrow\mathrm{S}}$ as a function of microwave frequency and initialization detuning $\epsilon$, defined as the detuning between the $(1,3)$ and $(0,4)$ charge configurations. At negative detuning, the spectrum shows four resonances corresponding to conditional-rotation (CROT) transitions of the two qubits under finite exchange ~\cite{noiri_fast_2022, Wu2025-highfidelity}. The exchange grows as the system approaches $\epsilon=0$, where wave-function overlap and charge hybridization increase. Inside the Pauli-blockade readout window, approximately $\epsilon=0$ to \SI{0.25}{meV}, the spectrum is dominated by exchange and charge hybridization. Deep in the $(0,4)$ region, the resonances reduce to the two degenerate triplet transitions $\ket{(0,4)T_{0}}\leftrightarrow\ket{(0,4)T_{-}}$ and $\ket{(0,4)T_{0}}\leftrightarrow\ket{(0,4)T_{+}}$, which are used for microwave initialization.
 
The transition positions are reproduced by the eight-state spin-charge Hamiltonian described in Methods. The model includes the relevant Zeeman energies, singlet and triplet charge branches, and effective tunnel couplings between states with the appropriate spin-charge character, with a singlet tunnel coupling of order $t_c\simeq\SI{0.5}{GHz}$ and a singlet-triplet splitting \(\Delta_\mathrm{ST}\simeq\SI{0.25}{meV}\). The calculated spectrum in Fig.~\ref{fig:fig2}e explains the measured resonance positions in Fig.~\ref{fig:fig2}d: the large qubit splittings, of order \SI{11.58}{GHz}, are set by the external field and micromagnet magnetization, while exchange and charge hybridization generate the smaller detuning-dependent splittings. 
Based on this spectrum, we choose an initialization point deep in the $(0,4)$ triplet manifold for the main protocol (green triangle in Fig.~\ref{fig:fig2}d). At this point, the polarized-to-unpolarized triplet resonances are nearly degenerate, while the $\ket{(0,4)S}$ branch is spectrally separated from the driven transition. This suppresses unwanted excitation out of $\ket{(0,4)S}$ and makes the microwave step robust in the present device. 

An alternative implementation of microwave initialization operates within (or near) the Pauli-spin-blockade window and directly addresses the exchange-split blocked-state resonances of the $(1,3)$-like even branches, for example, near $\epsilon_\mathrm{M}$ indicated by the red triangle in Fig.~\ref{fig:fig2}d. This in-window approach is demonstrated in Extended Data Fig.~\ref{fig_ext:fig1} using chirped microwave pulses. In the present device, however, it is less favorable because multiple resonances must be addressed and microwave excitation close to the charge transition also drives population loss from $\ket{(0,4)S}$. The measurements show that this singlet depopulation increases strongly with microwave amplitude and extends over a broad detuning range, limiting repeated initialization in the present device.

\subsection{Microwave transfer and pumping}
\begin{figure}[H]
 \centering
 \captionsetup{font=footnotesize,skip=0pt,width=1\linewidth}
 \includegraphics[width=1\textwidth]{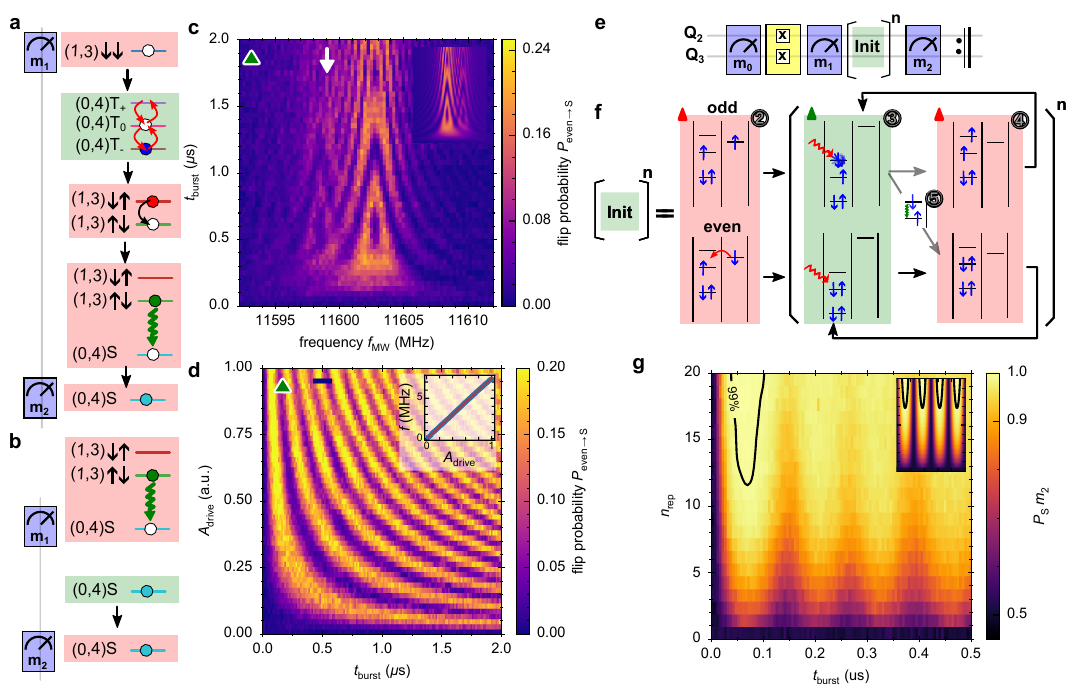}
 \caption{\textbf{Microwave-driven triplet transfer and repeated initialization.}
 \textbf{a}, Simplified sequence for an initially polarized even state, illustrated for $\ket{\downarrow\downarrow}$.
 \textbf{b}, Corresponding sequence for an initially odd state, illustrated for $\ket{\uparrow\downarrow}$.
 \textbf{c}, Resonant triplet drive in the $(0,4)$ manifold. The chevron-like pattern reflects coherent transfer between polarized triplets and $\ket{(0,4)T_{0}}$, with a maximum coherent transfer of 50\% from one polarized triplet at common resonance because of a dark state. A weak additional resonance is attributed to a valley-related state.
 \textbf{d}, Rabi frequency of the triplet transition as a function of microwave drive amplitude, showing the expected linear EDSR scaling.
 \textbf{e}, Randomizing circuit based on $X/2$ rotations of Q$_2$ and Q$_3$.
 \textbf{f}, Repetition of the initialization cycle.
 \textbf{g}, Singlet probability \(P_S\) versus repetition number \(n_\mathrm{rep}\) for a randomized input, reaching 99\% after 11 repetitions. \(P_S\) denotes the measured \((0,4)\) charge-outcome probability, assigned to the odd-parity outcome in Pauli-spin-blockade readout.}
 \label{fig:fig3}
\end{figure}

Figure~\ref{fig:fig3} characterizes microwave transfer and repeated pumping, with sequences for representative even and odd inputs shown in Fig.~\ref{fig:fig3}a,b.

Driving one of the triplet resonances identified in Fig.~\ref{fig:fig2} produces the chevron-like response in Fig.~\ref{fig:fig3}c. The suppressed conversion at its center indicates the presence of a dark state and is reproduced by the driven three-state triplet model shown in the inset (see also Methods and Extended Data Fig.~\ref{fig_ext:fig2}). The model uses the basis $\{\ket{(0,4)T_{-}},\ket{(0,4)T_{0}},\ket{(0,4)T_{+}}\}$. Because $\ket{(0,4)T_{0}}$ couples to both polarized triplets, it is useful to define a bright state and a dark state, $\ket{B}=(\ket{(0,4)T_{-}}+\ket{(0,4)T_{+}})/\sqrt{2}$ and $\ket{D}=(\ket{(0,4)T_{-}}-\ket{(0,4)T_{+}})/\sqrt{2}$. Only $\ket{B}$ couples directly to $\ket{(0,4)T_{0}}$, whereas population in $\ket{D}$ remains there during a pulse at the common resonance of the two equally driven transitions~\cite{MorrisShore1983}. A single polarized triplet contains equal bright and dark components, so at most half of the population can be transferred into $\ket{(0,4)T_{0}}$ by a coherent resonant pulse. Away from common resonance, detuning mixes the bright and dark states. For the randomized two-spin input used here, half of the ensemble is initially in the polarized even states, so the maximum additional conversion in a single ideal microwave cycle is $0.5\times0.5=0.25$.

The weaker resonance in Fig.~\ref{fig:fig3}c (white arrow) is attributed to an excited valley state in Q$_3$. The main-resonance conversion reaches 0.18, below the ideal mixed-ensemble limit of 0.25 (Extended Data Fig.~\ref{fig_ext:fig2}).

The microwave transfer rate is controlled by the drive amplitude. As shown in Fig.~\ref{fig:fig3}d, the resonant oscillation frequency is linear in the microwave amplitude, as expected for EDSR, and reaches about \SI{8}{MHz} at the largest drive. At the stronger drive used for the mixing measurements, the model predicts transfer close to the 50\% dark-state limit for both valley branches, supporting efficient population transfer into the $T_0$-like pathway (Extended Data Fig.~\ref{fig_ext:fig2}d,e).

Repetition overcomes the single-cycle transfer limit: the observed accumulation indicates that inter-cycle evolution makes residual blocked population addressable again. For the randomized input in Fig.~\ref{fig:fig3}e,g, $P_S$ reaches 99\% after 11 repetitions with \SI{65}{ns} bursts. Here $P_S$ is the singlet-associated $(0,4)$ charge-outcome probability assigned to odd parity. Each cycle combines microwave transfer of blocked population with probability $p_\mathrm{MW}$ and subsequent singlet conversion with conditional probability $q$. The repetition dependence is summarized by $P_S(n)=1-0.5(1-p_\mathrm{eff})^n$, with $p_\mathrm{eff}=0.343$, the net per-cycle gain combining these processes.

\subsection{Initialization into the target \texorpdfstring{$\ket{\uparrow\downarrow}$}{up-down} state}
\begin{figure}[H]
 \centering
 \captionsetup{font=footnotesize,skip=0pt,width=1\linewidth}
 \includegraphics[width=1\textwidth]{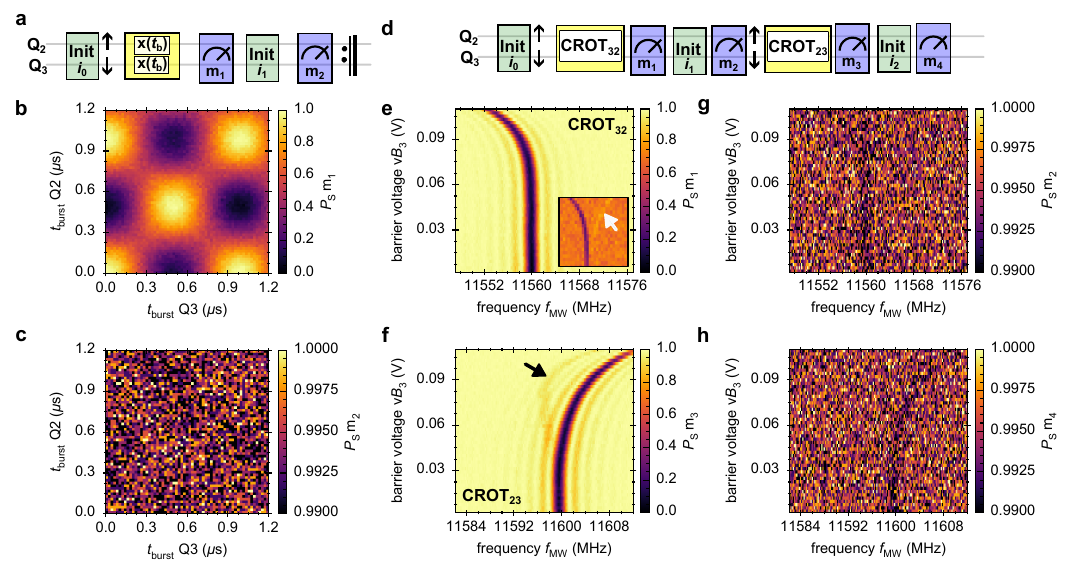}
 \caption{\textbf{Verification of initialization into the target operational state.}
 \textbf{a}, Sequence used to vary the prepared two-spin input state and then apply the initialization protocol.
 \textbf{b}, Singlet probability before the final initialization step, showing the expected checkerboard pattern from coherent Q$_2$ and Q$_3$ rotations.
\textbf{c} Singlet probability after initialization, with a median $P_\mathrm{S}=99.4\%$ and a standard deviation of $0.37\%$ across the sampled preparation states. \textbf{d} Sequence for exchange spectroscopy after initialization.
 \textbf{e} Conditional-rotation spectroscopy for $\mathrm{CROT}_{32}$, showing one dominant exchange branch.
 \textbf{f} Conditional-rotation spectroscopy for $\mathrm{CROT}_{23}$, also showing one dominant branch. The additional weak feature is attributed to a valley-related state.
 \textbf{g,h} Corresponding initialization readouts showing high singlet probability $P_{\mathrm{S}}$ after the reset step.}
 \label{fig:fig4}
\end{figure}

The measurements above demonstrate pumping into the $(0,4)$ charge configuration.
Charge sensing alone, however, does not distinguish $\ket{(0,4)S}$ from other $(0,4)$ states. 
Figure~\ref{fig:fig4} therefore verifies the initialized state in two steps.
First, we show that the initialization scheme prepares a singlet charge state that maps back to a coherent odd spin state at the operation point.
Second, we use exchange spectroscopy to identify the target $\ket{\uparrow\downarrow}$ state.

For the first test, we vary the prepared input state using independent rotations of the two spins after an initialization step $i_0$ in the sequence of Fig.~\ref{fig:fig4}a.
Microwave bursts on Q$_2$ and Q$_3$ rotate the two spins, and the singlet probability measured at $m_1$ forms the expected checkerboard pattern as a function of the two rotation angles (Fig.~\ref{fig:fig4}b).
This demonstrates coherent control starting from the state produced by the initialization step, consistent with mapping $\ket{(0,4)S}$ back to an odd spin state at the operation point.
The intermediate measurement $m_1$ characterizes the prepared inputs and is not required for the initialization protocol. After applying the second initialization step $i_1$, following the measurement $m_1$, the median singlet probability measured at $m_2$ is $99.4\%$, with a standard deviation of 0.37\% across the sampled preparation settings (Fig.~\ref{fig:fig4}c). The observed spread is consistent with finite sampling of 500 shots per point.
Together with the exchange-spectroscopy branch verification below, this demonstrates high-probability initialization across the tested input ensemble into the target operational state.

The same high measured singlet probability in Fig.~\ref{fig:fig4}c could, in principle, also arise if the protocol initialized into a mixture of the two odd states.
We therefore use exchange spectroscopy to identify the two-qubit state after initialization.
During operation, the virtualized interdot barrier gate $vB_3$ is varied to enable exchange.
A well-initialized $\ket{\uparrow\downarrow}$ state should show a single exchange branch in conditional-rotation spectroscopy.
A mixture of $\ket{\uparrow\downarrow}$ and $\ket{\downarrow\uparrow}$ would instead show a second branch associated with the opposite odd state, as shown in the inset of Fig.~\ref{fig:fig4}e.
Here $\mathrm{CROT}_{ij}$ denotes a conditional rotation where Q$_i$ is the control qubit and Q$_j$ is the driven qubit.

The data show one dominant exchange branch in both $\mathrm{CROT}_{32}$ (Fig.~\ref{fig:fig4}e) and $\mathrm{CROT}_{23}$ (Fig.~\ref{fig:fig4}f), with no resolved opposite-state branch. Fits that explicitly include this branch return amplitudes consistent with zero across the resolved Q$_2$ range. For each resolved Q$_2$ spectrum, any hidden opposite-state resonance contributes less than 1.08\% to the measured excitation probability at 95\% confidence (Extended Data Fig.~\ref{fig_ext:fig3}). Together with the high measured singlet outcome, this provides strong evidence for high-fidelity initialization into the target operational state. An additional feature in the $\mathrm{CROT}_{23}$ data (black arrow in Fig.~\ref{fig:fig4}f) follows a distinct trajectory and is attributed to the valley-related resonance also observed in Fig.~\ref{fig:fig3}c, rather than to the opposite odd-parity branch.

\subsection{Rate scale of blockade lifting}
\begin{figure}[H]
\centering
\captionsetup{font=footnotesize,skip=0pt,width=1\linewidth}
\includegraphics[width=1\textwidth]{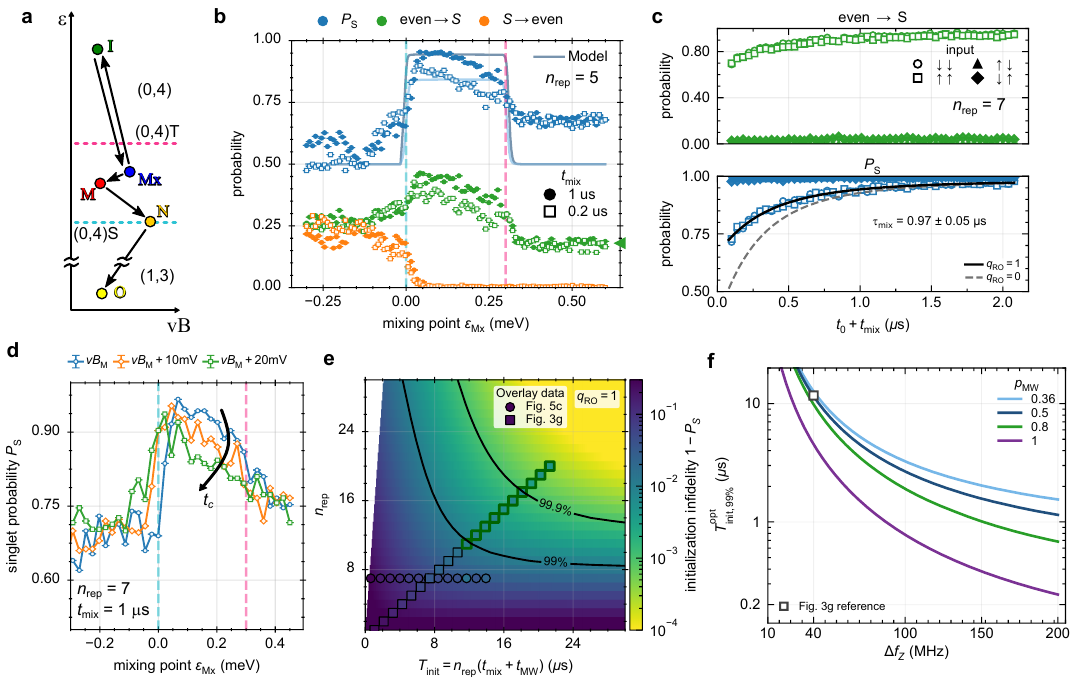}
\caption{\textbf{Mixing point and rate scale of the initialization protocol.}
\textbf{a} Initialization sequence with a mixing point MX chosen independently of the measurement point M. The dashed lines indicate the $(0,4)S$ and $(0,4)T$ transitions.
\textbf{b} Measured singlet probability for $n_\mathrm{rep}=5$ as a function of MX detuning. In these measurements, $vB_\mathrm{MX}$ is equivalent to $vB_\mathrm{M}$, the barrier voltage condition at the measurement point M.
\textbf{c} Singlet probability versus mixing dwell for four inputs after seven repetitions. Curves show the joint re-excitation fit to the even inputs with complete readout-assisted conversion ($q_\mathrm{RO}=1$, solid) and the same parameters without this conversion ($q_\mathrm{RO}=0$, dashed). Here $t_0=\SI{80}{ns}$ accounts for the fixed pulse timing, including ramps. Repetition gives a roughly $0.5~\mu\mathrm{s}$ rise despite a single-cycle constant $\tau_\mathrm{mix}=0.97\pm0.05~\mu\mathrm{s}$.
\textbf{d} Dependence on the interdot-barrier setting. Increasing the barrier voltage relative to the measurement point M, $vB_\mathrm{M}$, increases the effective tunnel coupling and lowers the conversion efficiency from the $T_0$-like pathway to $\ket{(0,4)S}$, consistent with suppressed blockade lifting at larger $t_c$ in this regime.
\textbf{e} Infidelity from the model in \textbf{c} with $q_\mathrm{RO}=1$, overlaid with $\ket{\downarrow\downarrow}$ data from \textbf{c} (circles) and independent \SI{65}{ns}-burst repetition data from Fig.~\ref{fig:fig3}g (squares, not fitted). The plotted time includes only microwave bursts and mixing dwells, and the probabilities include readout-assisted conversion.
\textbf{f} Projected minimum $T_\mathrm{init}$ for $P_S=0.99$, optimized over repetition number. The $p_\mathrm{MW}=0.36$, 0.5, 0.8, and 1 scenarios represent the low-drive reference from Fig.~\ref{fig:fig3}c, the resonant dark-state limit approached in Extended Data Fig.~\ref{fig_ext:fig2}d,e, higher in-window transfer motivated by Extended Data Fig.~\ref{fig_ext:fig1} without singlet depopulation, and ideal transfer. Methods and Extended Data Fig.~\ref{fig_ext:fig4} specify the model and gradient scaling. For comparison, the open square marker shows the unoptimized randomized-input experiment from Fig.~\ref{fig:fig3}g, with 11 repetitions and a \SI{1}{\micro\second} dwell.
}
\label{fig:fig5}
\end{figure}

The microwave pulse transfers blocked population into the $T_0$-like pathway within tens of nanoseconds, but initialization also requires its conversion to $\ket{(0,4)S}$. Figure~\ref{fig:fig5} examines where this conversion is most efficient, how long it takes and how it limits the overall initialization time. We introduce a mixing point MX independently of the measurement point M, allowing the conversion dwell to be adjusted without changing the readout conditions (Fig.~\ref{fig:fig5}a).

We first vary the MX detuning for five repetitions with a \SI{1}{\micro\second} dwell per cycle (Fig.~\ref{fig:fig5}b). Conversion is strongest near the singlet charge anticrossing, where singlet-triplet mixing and charge hybridization connect the $T_0$-like state to $\ket{(0,4)S}$. Moving too close to the anticrossing also increases unwanted even-parity outcomes for initially odd inputs. We therefore choose MX in the middle of the Pauli-blockade window, balancing efficient conversion against singlet loss. The residual even-to-singlet signal of about 0.18 with MX deep in $(0,4)$ is consistent with conversion outside the mixing dwell, including during the final readout at M.

At this mixing point, increasing $t_\mathrm{mix}$ raises $P_S$ for initially even inputs, while initially odd inputs already give high $P_S$ (Fig.~\ref{fig:fig5}c). After seven repetitions, even inputs yield $P_S\approx70\%$ at the shortest dwell because conversion also occurs along the fixed pulse trajectory and during readout. To extract the conversion time per cycle, we fit the two even-input traces with the re-excitation model, which allows later microwave bursts to return unconverted $T_0$-like population to the blocked states (Methods). The fit gives $p_\mathrm{MW}\simeq0.43$ and an effective single-cycle conversion constant $\tau_\mathrm{mix}=0.97\pm0.05~\mu\mathrm{s}$. Accumulation over seven repetitions produces the faster observed rise over roughly $0.5~\mu\mathrm{s}$ of dwell per cycle.

Increasing the tunnel coupling reduces conversion (Fig.~\ref{fig:fig5}d), consistent with the charge-dephasing-assisted scaling $\Gamma_\mathrm{blockade}\propto(\Delta f_Z/t_c)^2$~\cite{Seedhouse2021}. This identifies the subsequent blockade-lifting dynamics as the principal time scale, rather than the microwave burst.

Figure~\ref{fig:fig5}e compares the predicted infidelity with the mixing data and the independent repetition measurements from Fig.~\ref{fig:fig3}g. At large repetition numbers, the independent data approach a finite infidelity floor, limiting the benefit of further cycles. The model allows continued accumulation in the singlet and does not capture this saturation.

Finally, Fig.~\ref{fig:fig5}f projects the minimum microwave-plus-mixing time $T_\mathrm{init}=n_\mathrm{rep}(t_\mathrm{mix}+t_\mathrm{MW})$ required to reach $P_S=0.99$. We use the re-excitation model with readout-assisted and dwell-independent conversion removed, and optimize the repetition number for \SI{65}{ns} bursts (Methods and Extended Data Fig.~\ref{fig_ext:fig4}). The projections identify increased Zeeman-frequency difference and improved microwave transfer as a route to sub-microsecond initialization.

\section{Discussion}
The experiments establish reservoir- and measurement-free microwave initialization of a silicon spin-qubit pair. Repeated microwave and baseband pulses accumulate the tested input states in $\ket{(0,4)S}$, which is then mapped to the operational state. Parity readout verifies the high singlet-associated charge-outcome probability, and exchange spectroscopy identifies the target $\ket{\uparrow\downarrow}$ state.

Our measurements identify two central ingredients of the reset process. First, the microwave response shows that the deep-$(0,4)$ transfer step is dark-state limited, so one coherent burst cannot transfer all polarized-triplet population into the $T_0$-like pathway. Second, mixing-time measurements show that the subsequent blockade-lifting step sets the main time scale of the protocol. Repetition overcomes the single-cycle microwave-transfer limit because population that reaches $\ket{(0,4)S}$ is no longer resonantly driven by subsequent microwave pulses, while remaining blocked population is recycled through later attempts.

The effective single-cycle conversion time is approximately \SI{1}{\micro\second}. At comparable tunnel coupling, the expected $(\Delta f_Z)^2$ rate scaling and improved microwave transfer offer a route to sub-microsecond initialization (Fig.~\ref{fig:fig5}f).

Higher single-cycle transfer could be obtained by driving $(1,3)$-like blocked states within or near the Pauli-blockade window, avoiding the deep-$(0,4)$ triplet dark-state limit. Resolved exchange splitting is not essential: in the weak-exchange limit, a pulse on one qubit can address both $\ket{\downarrow\downarrow}\rightarrow\ket{\uparrow\downarrow}$ and $\ket{\uparrow\uparrow}\rightarrow\ket{\downarrow\uparrow}$ at the same frequency. Efficient initialization would still require subsequent singlet conversion and suppression of excitation out of $\ket{(0,4)S}$. In the present device, in-window driving was less robust because it also depopulated the singlet (Extended Data Fig.~\ref{fig_ext:fig1}). Larger singlet-triplet and valley-state separations, together with pulse optimization, could improve this selectivity.

The demonstrated protocol enables reset of an already loaded spin pair using a fixed microwave and baseband sequence, allowing initialization sites to be placed independently of reservoirs and charge sensors. Because it relies on two-spin singlet-triplet physics, the principle can be adapted to other semiconductor platforms with suitable Pauli-blockade transitions, including SiMOS electron-spin and hole-spin systems. By removing the need to colocate every initialization location with reservoir or measurement infrastructure, this approach addresses an architectural constraint that becomes increasingly important as processor size grows.


\section{Methods}
\subsection{Device and qubit operation}
The experiment uses a three-qubit Si/SiGe device fabricated by Intel. The initialization protocol acts on Q$_2$ and Q$_3$, while a nearby charge sensor (CS in Fig.~\ref{fig:fig1}b) detects the charge state of the Q$_2$-Q$_3$ double dot using radio-frequency reflectometry with a standard NbTiN inductor chip. Micromagnets provide a magnetic-field gradient for qubit addressability and electrically driven spin resonance. The measurements reported here use Q$_2$ and Q$_3$ at an external magnetic field of \SI{0.35}{T}, where their EDSR frequencies differ by \SI{44.8}{MHz}. The two qubits are operated at point O, in the center of the $(1,3)$ charge region, where the exchange interaction is well below \SI{100}{kHz} and the product-state basis gives an accurate description of the separated spins.

 \subsection{Parity readout}
Readout uses Pauli spin blockade at the $(1,3)$-$(0,4)$ transition. At point N, the interdot barrier is opened to increase exchange and improve parity discrimination. Odd spin-parity states contain singlet-like components that hybridize with $\ket{(0,4)S}$ and relax to the $(0,4)$ singlet charge state in the blockade window. Even states remain blockaded in a $(1,3)$-like configuration. The charge state is then measured at M with a typical integration time of \SI{20}{\micro\second}.

The experiments use repeated charge measurements denoted $m_0$, $m_1$, $m_2$ and so on. For the uninitialized EDSR measurements in Fig.~\ref{fig:fig1}, a microwave pulse is inserted between two parity measurements. The charge-signal difference, shown in Fig.~\ref{fig:fig1}f, is obtained from the difference between the two sensor records and separates odd-to-even, even-to-odd and no-change events. This analysis gives access to the direction of the spin-parity transition, rather than only to the total spin-flip probability.

\subsection{Driven triplet-transfer model}
The microwave-transfer data in Fig.~\ref{fig:fig3}c and Extended Data Fig.~\ref{fig_ext:fig2} are described with a driven three-state model in the $(0,4)$ triplet manifold. In the rotating frame and within the rotating-wave approximation, we use the basis
\[
\left(
\ket{(0,4)T_{-}},
\ket{(0,4)T_{0}},
\ket{(0,4)T_{+}}
\right)
\]
and write
\begin{equation}
\frac{H_\mathrm{trip}}{h}
=
\begin{pmatrix}
+\delta_{-} & \Omega_{-}/2 & 0 \\
\Omega_{-}/2 & 0 & \Omega_{+}/2 \\
0 & \Omega_{+}/2 & -\delta_{+}
\end{pmatrix}.
\label{eq:triplet_drive_model}
\end{equation}
Here $\delta_{\pm}=f_\mathrm{MW}-f_{T_0T_{\pm}}$ are the microwave detunings from the two polarized-triplet transitions, and $\Omega_{\pm}$ are the corresponding drive rates. The transition frequencies are defined as positive energy gaps, $f_{T_0T_{-}}=(E_{T_0}-E_{T_{-}})/h$ and $f_{T_0T_{+}}=(E_{T_{+}}-E_{T_0})/h$. 
At common resonance, $\delta_-=\delta_+=0$, and for equal drive amplitudes, only the symmetric superposition of the polarized triplets couples to $\ket{(0,4)T_0}$ while the antisymmetric superposition remains dark. A polarized triplet contains equal components of these superpositions, limiting resonant single-pulse transfer to 50\%. Away from common resonance, detuning mixes the bright and dark states.

The chevron in Extended Data Fig.~\ref{fig_ext:fig2} is fitted by calculating the microwave-driven population transfer generated by Eq.~\ref{eq:triplet_drive_model}. The measured even-to-singlet signal is taken proportional to the transferred $\ket{(0,4)T_{0}}$ population after the subsequent baseband conversion step. A weaker additional resonance is included phenomenologically with an independent amplitude and resonance frequency, and is attributed to a valley-related excited state. The fit reproduces the measured conversion scale of approximately 0.18 at the main-resonance frequency for a mixed input ensemble. This value includes a contribution from the weaker valley-related resonance.

For the stronger-drive calculation, we fix the fitted resonance frequencies (\SI{11602.7735}{MHz} and \SI{11598.4956}{MHz}), signal weights and background, and scale both couplings by 4.44, following the increase in $f_\mathrm{Rabi}=\sqrt{2}\Omega$ from 1.68 to \SI{7.48}{MHz} for equal couplings $\Omega_\pm=\Omega$. At \SI{11602.768}{MHz} and \SI{70}{ns}, the conditional main- and valley-branch transfers are 0.497 and 0.477, and the weighted ensemble signal is 0.217. This assumes coherent, independently additive branch responses. The fitted weights are not valley populations, so this calculation does not independently calibrate $p_\mathrm{MW}$ for the mixing scans.

\subsection{Eight-state Hamiltonian model}
An effective eight-state spin-charge Hamiltonian reproduces the resonance positions and anticrossing structure in Fig.~\ref{fig:fig2}d,e.

In the product-state basis
\[
\mathcal{B}=
\left(
\ket{\downarrow\downarrow},
\ket{\uparrow\downarrow},
\ket{\downarrow\uparrow},
\ket{\uparrow\uparrow},
\ket{(0,4)S},
\ket{(0,4)T_{-}},
\ket{(0,4)T_{0}},
\ket{(0,4)T_{+}}
\right),
\]
where the first and second arrows refer to the left and right spin, respectively, the Hamiltonian used for the resonance overlay is written in frequency units as
\begin{equation}
\resizebox{\textwidth}{!}{$
\frac{H}{h}=
\begin{pmatrix}
-\frac{f_{Z,L}+f_{Z,R}}{2} & 0 & 0 & 0 & 0 & t_T & 0 & 0 \\
0 & \frac{f_{Z,L}-f_{Z,R}}{2} & 0 & 0 & t_S & 0 & t_{UD,T_0} & 0 \\
0 & 0 & \frac{-f_{Z,L}+f_{Z,R}}{2} & 0 & -t_S & 0 & t_{DU,T_0} & 0 \\
0 & 0 & 0 & \frac{f_{Z,L}+f_{Z,R}}{2} & 0 & 0 & 0 & t_T \\
0 & t_S & -t_S & 0 & \Delta-\epsilon & 0 & 0 & 0 \\
t_T & 0 & 0 & 0 & 0 & \Delta-\epsilon+\Delta_\mathrm{ST}-f_{Z,R} & 0 & 0 \\
0 & t_{UD,T_0} & t_{DU,T_0} & 0 & 0 & 0 & \Delta-\epsilon+\Delta_\mathrm{ST} & 0 \\
0 & 0 & 0 & t_T & 0 & 0 & 0 & \Delta-\epsilon+\Delta_\mathrm{ST}+f_{Z,R}
\end{pmatrix}
$}
\label{eq:eight_state_hamiltonian}
\end{equation}

The offset $\Delta$ defines the detuning origin. With $\Delta=0$, positive $\epsilon$ is defined towards the $(0,4)$ charge state.
Here \(f_{Z,L}\) and \(f_{Z,R}\) are the Zeeman frequencies of the two spins, with \(\Delta f_Z=f_{Z,R}-f_{Z,L}\). The \((0,4)\) polarized triplet energies use \(f_{Z,R}\), corresponding to the Zeeman scale of the high-occupancy dot. The signs of the singlet couplings reflect the convention
\(\ket{(1,3)S}=(\ket{\uparrow\downarrow}-\ket{\downarrow\uparrow})/\sqrt{2}\), whereas the
\(\ket{(1,3)T_0}\)-like couplings have the same sign.

For the overlay in Fig.~\ref{fig:fig2}d, we use
\[
f_{Z,L}=\SI{11.561}{GHz},\quad
f_{Z,R}=\SI{11.601}{GHz},\quad
\Delta f_Z=\SI{40}{MHz}.
\]

The value for $\Delta f_Z$ differs slightly from the operation-point qubit-energy separation presented in Fig.~\ref{fig:fig1}c, because the spectrum is measured at different voltage conditions.

\[
\Delta=0,\quad
\Delta_\mathrm{ST}\simeq\SI{0.25}{meV},
\]
and the effective tunnel couplings
\[
t_c=\SI{0.5}{GHz},\quad
t_S=\frac{t_c}{\sqrt{2}},\quad
t_T=\SI{0.35}{GHz},\quad
t_{UD,T_0}=t_{DU,T_0}=\frac{\SI{0.3}{GHz}}{\sqrt{2}}.
\]
Here $t_c$ denotes the singlet--singlet tunnel coupling, giving the product-state matrix elements $t_S=t_c/\sqrt{2}$.
In Eqs.~\ref{eq:eight_state_hamiltonian} and~\ref{eq:mixing_hamiltonian}, detunings and splittings are expressed in frequency units. Quoted energy scales such as $\Delta_\mathrm{ST}$ and the detuning axis use multiplication by $h$. The detuning axis is calibrated using $\alpha_\epsilon\simeq\SI{0.15}{meV/mV}$, corresponding to opposite-sign changes of the two relevant virtual plungers.
The spectrum parameters were adjusted manually to match the observed resonance positions and anticrossings. Offsets from residual spin-spin exchange in $(1,3)$ are neglected because the exchange at the operation point is negligible on the scale of the microwave spectrum. The detuning-dependent splittings in the overlay arise from coupling to the singlet and triplet charge branches.

The Hamiltonian in Eq.~\ref{eq:eight_state_hamiltonian} describes the coherent spin-charge spectrum in Fig.~\ref{fig:fig2}d,e. The irreversible part of the initialization, namely conversion from a $T_0$-like state to the singlet charge state, is described separately by a minimal blockade-lifting model \cite{Seedhouse2021}. In the basis

\[
\left(
\ket{(0,4)S},
\ket{(1,3)S},
\ket{(1,3)T_{0}}
\right),
\]
we use
\begin{equation}
\frac{H_\mathrm{mix}}{h}
=
\begin{pmatrix}
-\epsilon & t_c & 0 \\
t_c & 0 & -\Delta f_Z/2 \\
0 & -\Delta f_Z/2 & 0
\end{pmatrix}.
\label{eq:mixing_hamiltonian}
\end{equation}
Charge dephasing can assist blockade lifting along the $\ket{(1,3)T_0}\rightarrow\ket{(1,3)S}\rightarrow\ket{(0,4)S}$ pathway, while relaxation provides irreversible capture in the singlet ground state. In the charge-dephasing-limited regime, $\Gamma_\mathrm{blockade}\propto(\Delta f_Z/t_c)^2$, motivating the barrier interpretation and gradient projections. The population model below describes this net conversion through an effective single-cycle constant $\tau_\mathrm{mix}$.

\subsection{Mixing-time analysis}
The mixing-point experiment in Fig.~\ref{fig:fig5} varies the conversion dwell at fixed microwave settings. In the detuning scan, $n_\mathrm{rep}=5$ is fixed and MX is varied. In the time-domain experiment, $n_\mathrm{rep}=7$ is fixed and the mixing time $t_\mathrm{mix}$ at MX is varied for all four input states.

Let $t=t_\mathrm{mix}$ and $n=n_\mathrm{rep}$. In the simple retry model, each burst transfers blocked population to the convertible pathway with probability $p_\mathrm{MW}$, followed by singlet conversion
\begin{equation}
q_\mathrm{relax}(t)=1-(1-q_0)e^{-t/\tau_\mathrm{mix}}.
\label{eq:conditional_relaxation}
\end{equation}
Here $q_0$ is the effective trajectory contribution at zero dwell and $\tau_\mathrm{mix}$ the single-cycle constant. The singlet population is absorbing, but all remaining population is available for another attempt.

Additional conversion during the \SI{20}{\micro\second} verification readout gives
\begin{equation}
q_\mathrm{final}(t)=q_\mathrm{relax}(t)+[1-q_\mathrm{relax}(t)]q_\mathrm{RO},
\label{eq:final_conversion}
\end{equation}
where $q_\mathrm{RO}$ is the conversion probability for residual convertible population, not readout assignment fidelity. For initial blocked fraction $P_{B,0}$,
\begin{equation}
P_S^{(n)}(t)=1-P_{B,0}[1-p_\mathrm{MW}q_\mathrm{relax}(t)]^{n-1}
[1-p_\mathrm{MW}q_\mathrm{final}(t)],\qquad n\geq1.
\label{eq:repeated_conversion}
\end{equation}
The baseline is $P_S^{(0)}=1-P_{B,0}$, with $P_{B,0}=1$ for polarized inputs and 0.5 for an ideal randomized input. We assume complete final conversion, $q_\mathrm{RO}=1$. For $n\geq2$, $0<p_\mathrm{MW}<1$ and $q_0<1$, the dwell response contains rates $k/\tau_\mathrm{mix}$, $k=1,\ldots,n-1$, allowing an initial rise faster than $\tau_\mathrm{mix}$.

We jointly fit the two polarized-input traces by equal-weight least squares using unchanged charge thresholds (51 dwell settings per input, 1,000 shots per setting). Bounds are $0.15\leq p_\mathrm{MW}\leq0.5$, $0\leq q_0\leq1$ and $0.02~\mu\mathrm{s}\leq\tau_\mathrm{mix}\leq10~\mu\mathrm{s}$. The simple model gives $p_\mathrm{MW}=0.424\pm0.009$, $q_0=0.273\pm0.009$ and $\tau_\mathrm{mix}=0.85\pm0.05~\mu\mathrm{s}$. Quoted errors are residual-scaled local standard errors for correlated fit parameters, conditional on the preparation, conversion and assignment model but they exclude calibration and model uncertainty, including reverse transfer.

The low-drive scale $2\times0.18=0.36$ is not a matched mixing calibration because the drive conditions differ and the measured signal includes capture and overlapping resonances. We therefore fit $p_\mathrm{MW}$ independently for the mixing scans.

The re-excitation model tracks blocked, convertible and singlet populations $(B,T,S)$, initially $(P_{B,0},0,1-P_{B,0})$. Each burst applies
\[
B'=(1-p_\mathrm{MW})B+p_\mathrm{MW}T,\qquad
T'=p_\mathrm{MW}B+(1-p_\mathrm{MW})T,
\]
followed by
\[
S_\mathrm{next}=S+qT',\qquad
T_\mathrm{next}=(1-q)T',\qquad B_\mathrm{next}=B',
\]
with $q=q_\mathrm{relax}$ for the first $n-1$ cycles and $q=q_\mathrm{final}$ for the last. Equal forward and reverse transfer is a phenomenological assumption, and the singlet remains absorbing. For this symmetric re-excitation model, the same fit procedure with $n=7$ and $q_\mathrm{RO}=1$ gives $p_\mathrm{MW}=0.429\pm0.008$, $q_0=0.203\pm0.005$ and $\tau_\mathrm{mix}=0.97\pm0.05~\mu\mathrm{s}$. 
To test how the assumed reverse transfer affects the extracted conversion time, we refit the same two traces in Fig.~\ref{fig:fig5}c using a reverse-transfer probability of $2p_\mathrm{MW}$. This relation follows for equally driven, degenerate triplet transitions when coherence between triplet states is lost between cycles. The fit remains comparably good and gives $\tau_\mathrm{mix}\approx0.71~\mu\mathrm{s}$, showing the dependence of the extracted time on the transfer model.

Figure~\ref{fig:fig5}e uses the re-excitation fit with polarized input and $q_\mathrm{RO}=1$. Time is $T_\mathrm{init}=n_\mathrm{rep}(t_\mathrm{mix}+t_\mathrm{MW})$, with \SI{70}{ns} bursts for the mixing data and model map, and \SI{65}{ns} for the independent repetition data.

For Fig.~\ref{fig:fig5}f and Extended Data Fig.~\ref{fig_ext:fig4}, we assume polarized input, $q_0=q_\mathrm{RO}=0$, no singlet depopulation, \SI{65}{ns} bursts and $\tau_\mathrm{mix}=0.970~\mu\mathrm{s}(40~\mathrm{MHz}/\Delta f_Z)^2$. For each $p_\mathrm{MW}$ scenario and integer $n_\mathrm{rep}$, we find the shortest dwell reaching $P_S=0.99$. Fig.~\ref{fig:fig5}f minimizes $T_\mathrm{init}$ over $n_\mathrm{rep}$. The gradient scaling assumes unchanged mixing-point detuning,
tunnel coupling and charge dephasing. Both timing calculations exclude fixed transport overheads and verification readout. Measured probabilities can include additional conversion of residual $T_0$-like population during readout.


\section*{Use of artificial intelligence tools}
OpenAI ChatGPT and Codex were used to assist with language editing and manuscript preparation, and with the development and debugging of data-analysis and plotting code. The authors critically reviewed and verified all AI-assisted content and code.

\section*{Data Availability}
The data supporting the findings of this study are available in Zenodo at \url{https://doi.org/10.5281/zenodo.22806668}~\cite{Camenzind2026_mwinit_data}.

\section*{Code Availability}
Custom analysis code used to generate the results reported in this manuscript is available from the corresponding authors upon reasonable request.

\bibliographystyle{sn-standardnature_nourl_url}
\bibliography{bibfile}

\section*{Acknowledgements}
We thank the Intel team for providing the industrial Si/SiGe spin-qubit devices used in this work.
This work was supported by JST Moonshot R\&D Grant Number JPMJMS226B and JPMJMS256H. We acknowledge support from JSPS KAKENHI Grant Numbers 26K00642 (L.C.C.), 22H01160 (T.K.) and 23H05455 (K.T.), and JST PRESTO Grant Numbers JPMJPR2017 (T.N.) and JPMJPR23F8 (A.N.), and RIKEN Incentive Research Project Grant Number 202601080071 (L.C.C.).

\section*{Author Contributions}
L.C.C. and I.K.J. performed the experiment and analysed the data. A.N., K.T., T.N., and T.K. contributed to the data acquisition and discussed the results. L.C.C. wrote the manuscript with inputs from all co-authors. S.T. supervised the project.

\section*{Competing interests}
The authors declare no competing interests.

\section*{Extended Data}\label{extended_data}
\renewcommand{\figurename}{Extended data figure}
\renewcommand{\tablename}{Extended data table}
\setcounter{figure}{0} 

\begin{figure}[H]
\centering
\captionsetup{font=footnotesize,skip=0pt,width=1\linewidth}
\includegraphics[width=1\textwidth]{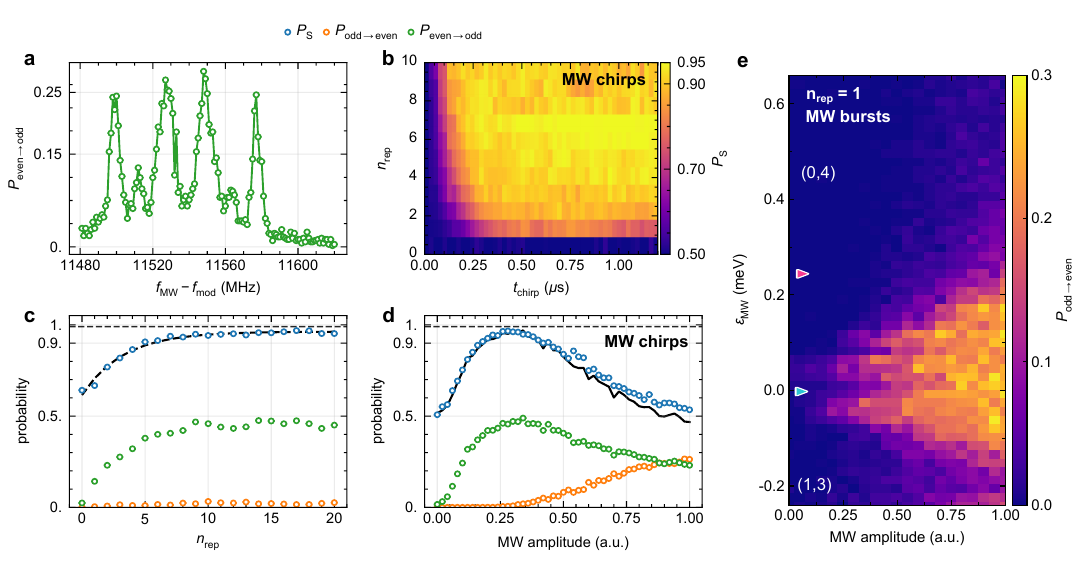}
\caption{\textbf{Initialization inside the Pauli-spin-blockade window.}
\textbf{a} Microwave spectrum measured inside the Pauli-spin-blockade window for a randomized two-spin input, showing several exchange-split resonances that can contribute to even-to-singlet conversion. The spectrum was acquired using a \SI{0.5}{\micro\second} chirped microwave burst with a rectangular amplitude envelope and a \SI{4}{MHz} chirp span. Even-to-singlet and even-to-odd components are extracted from the charge-signal difference, as described in Fig.~\ref{fig:fig1}.
\textbf{b} Measured singlet probability as a function of chirp duration and number of initialization repetitions. Sequential pulses, each of duration $t_\mathrm{chirp}$, sweep a \SI{4}{MHz} window around each selected resonance in \textbf{a}.
\textbf{c} Repetition dependence for the chirped initialization sequence. The singlet probability increases with the number of repetitions, but saturates below the target \(P_\mathrm{S}=99\%\). This limitation is consistent with a finite probability of unwanted singlet-to-even transitions (orange points).
\textbf{d} Microwave-amplitude dependence of initialization. Increasing the drive initially enhances even-to-singlet conversion, but stronger driving also increases singlet depopulation and reduces the final initialization probability. Similar limitations occur for sequential rectangular microwave bursts (\textbf{e}).
\textbf{e} Odd-to-even transition probability versus microwave amplitude and detuning for one initialization cycle using sequential rectangular microwave bursts at the selected resonance frequencies. The associated singlet depopulation is strongly amplitude- and detuning-dependent and broadens over a wider detuning range at larger drive amplitudes.
}
\label{fig_ext:fig1}
\end{figure}

\begin{figure}[H]
\centering
\captionsetup{font=footnotesize,skip=0pt,width=1\linewidth}
\includegraphics[width=1\textwidth]{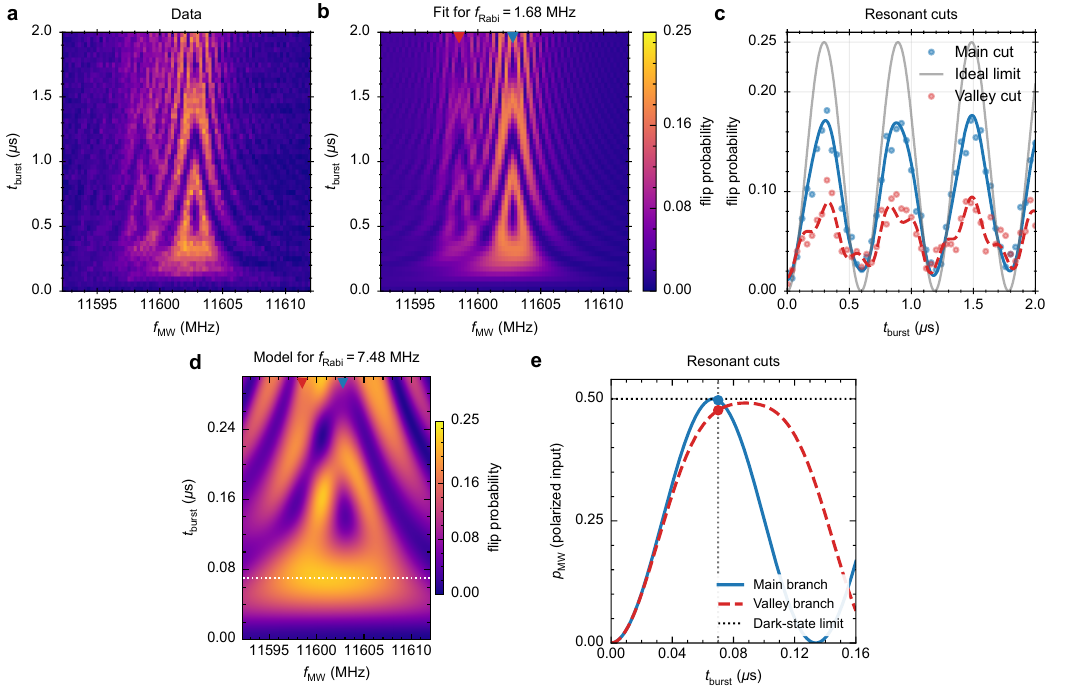}
\caption{\textbf{Three-state triplet model for the microwave-driven transfer.}
\textbf{a}, Measured microwave chevron for resonant transfer in the $(0,4)$ triplet manifold, plotted as even-to-singlet flip probability versus microwave frequency and burst duration.
\textbf{b}, Fit to the driven triplet model discussed in Methods, including the main triplet resonance and a weaker additional resonance attributed to a valley-related state.
\textbf{c}, Time-domain cuts through the main resonance and the weaker valley-related resonance. The cuts reach approximately 0.18 and 0.08. These overlapping signals are not calibrated branch populations. The suppressed conversion at the center of the chevron is reproduced by the dark-state-limited triplet model. The finite flip probability at nonzero $t_{\mathrm{burst}}$ indicates finite coupling to the $\ket{(0,4)T_{0}}$ state.
\textbf{d}, Calculated chevron using the fitted parameters from \textbf{b}, with the microwave drive scaled to $f_{\mathrm{Rabi}}=\SI{7.48}{MHz}$. The white dotted line marks $t_\mathrm{burst}=\SI{70}{ns}$, the burst duration used for the mixing measurements. The input ensemble is the same as in \textbf{a} and \textbf{b}, and the calculated flip probability approaches the mixed-ensemble limit of 0.25.
\textbf{e}, Calculated $p_\mathrm{MW}$ for polarized inputs on the main and valley branches at the same applied microwave frequency. At \SI{70}{ns}, $p_{\mathrm{MW}}=0.497$ and $0.477$, respectively. The black dotted line marks the dark-state limit of $0.5$.
}
\label{fig_ext:fig2}
\end{figure}

\begin{figure}[H]
\centering
\captionsetup{font=footnotesize,skip=0pt,width=1\linewidth}
\includegraphics[width=1\textwidth]{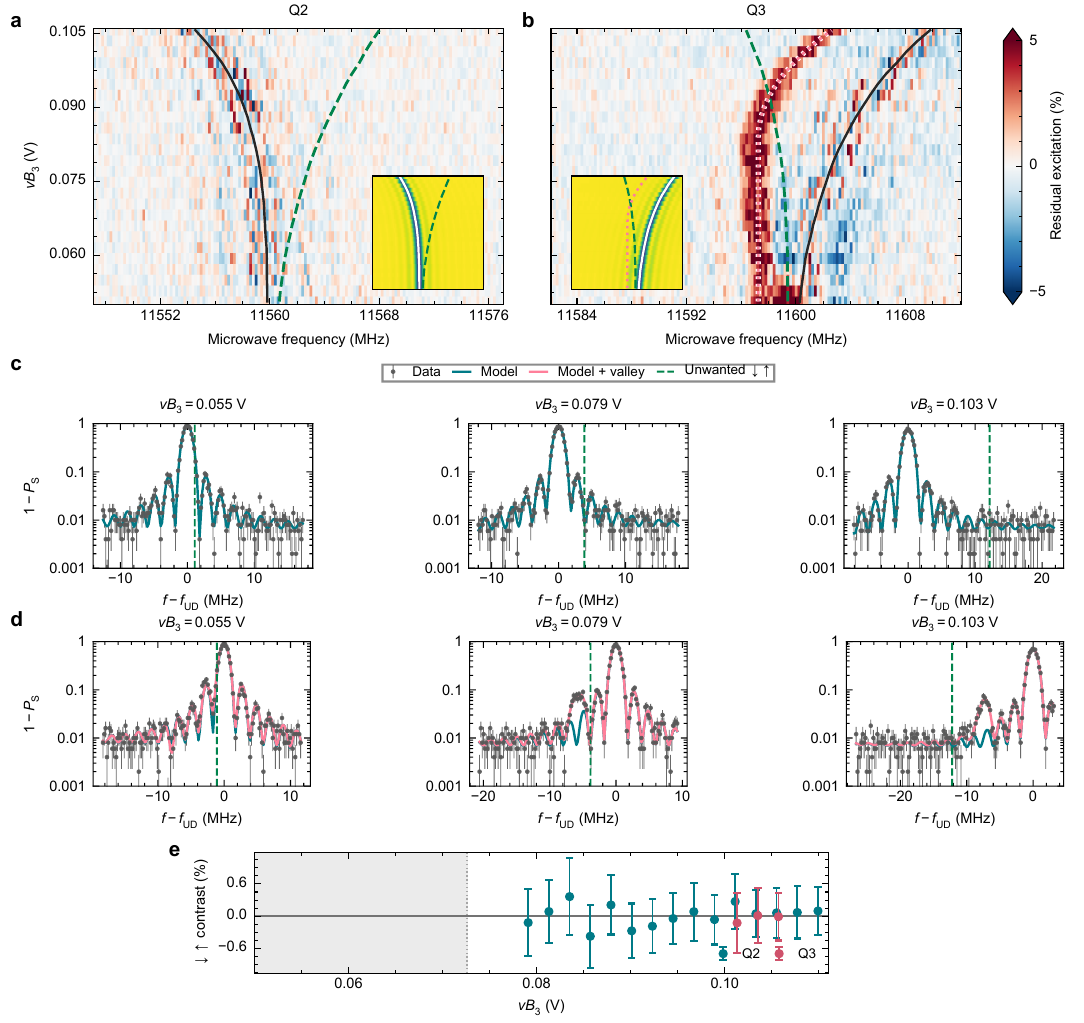}
\caption{\textbf{Exchange-resolved test of initialization into the target spin state.}
\textbf{a,b}, Residual $\mathrm{CROT}_{32}$ and $\mathrm{CROT}_{23}$ excitation after subtracting the fitted $\ket{\uparrow\downarrow}$ target response and smooth background. The solid black curve marks the fitted target resonance and the green dashed curve the predicted $\ket{\downarrow\uparrow}$ branch. No residual feature follows the predicted opposite-state trajectory. In \textbf{b}, the prominent residual instead follows the pink dotted trajectory of the known Q$_3$ valley-related resonance. The insets show the target-response model. The valley contribution is omitted from the Q$_3$ model and the pink dotted line is shown only as a positional guide.
\textbf{c,d}, Representative Q$_2$ and Q$_3$ excitation traces. The green dashed line marks the predicted opposite-state resonance. The Q$_2$ data are reproduced by the target-state response alone. For Q$_3$, blue shows the background and target contribution from the full fit. The pink curve additionally includes valley and opposite-state terms. No additional CROT branch is resolved.
\textbf{e}, Fitted opposite-branch contrast with 95\% bootstrap confidence intervals evaluated separately at each barrier voltage. The two spectra were fitted jointly, including the target and opposite-state resonances and the additional valley-related feature in Q$_3$. Confidence intervals were obtained by repeatedly generating and refitting synthetic data with the same shot noise expected for the measurements. The grey region indicates unresolved spin branches.
Across the resolved Q$_2$ range, no opposite-state resonance is observed, with individual upper bounds no larger than 1.08\% in excitation contrast.
}
\label{fig_ext:fig3}
\end{figure}

\begin{figure}[H]
\centering
\captionsetup{font=footnotesize,skip=0pt,width=1\linewidth}
\includegraphics[width=1\textwidth]{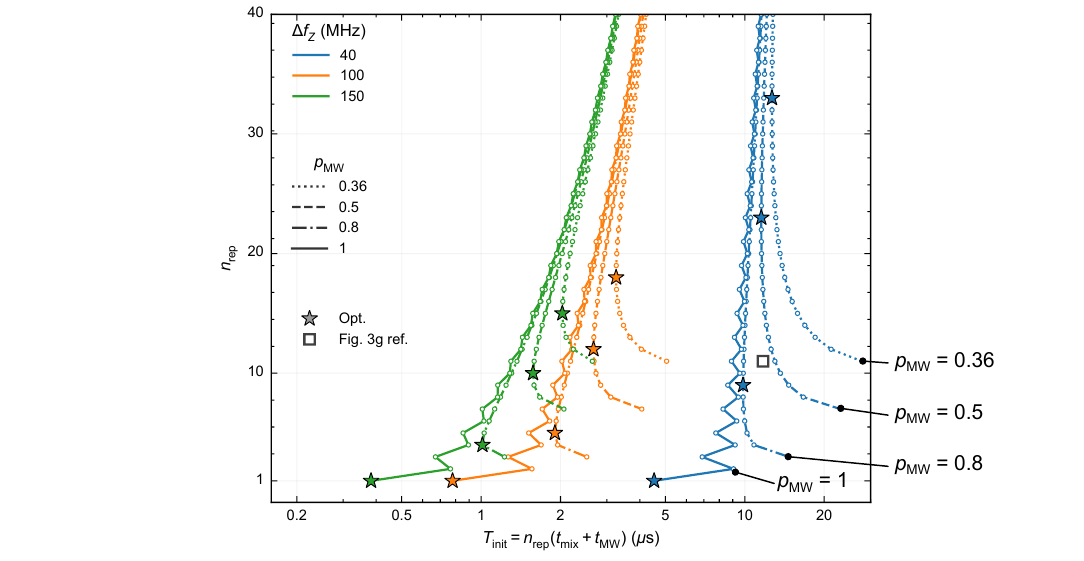}
\caption{
\textbf{Initialization time as a function of the number of repetitions.}
Calculated $P_\mathrm{S}=0.99$ contours for $\Delta f_Z=40$, 100 and \SI{150}{MHz}. Colors indicate $\Delta f_Z$, while line styles indicate the microwave-transfer probability $p_\mathrm{MW}$. At each integer repetition number $n_\mathrm{rep}$, $t_\mathrm{mix}$ is the minimum mixing dwell required to reach $P_\mathrm{S}=0.99$ in the re-excitation model. The initialization time is defined as in Fig.~\ref{fig:fig5}f, $T_\mathrm{init}=n_\mathrm{rep}(t_\mathrm{mix}+t_\mathrm{MW})$, with $t_\mathrm{MW}=\SI{65}{ns}$. Open circles mark the calculated solutions, and stars mark the minimum $T_\mathrm{init}$ for each $(\Delta f_Z,p_\mathrm{MW})$ pair. These optimum values are summarized in Fig.~\ref{fig:fig5}f. The calculations assume a polarized blocked input, no singlet depopulation, $q_0=q_\mathrm{RO}=0$, and
$\tau_\mathrm{mix}=0.970~\mu\mathrm{s}(40~\mathrm{MHz}/\Delta f_Z)^2$, using the single-cycle constant from the free-transfer re-excitation fit (Methods).
The four $p_\mathrm{MW}$ scenarios are defined in Fig.~\ref{fig:fig5}f. These are not independent transfer calibrations for the mixing scans.
For $p_\mathrm{MW}=1$, each microwave burst completely transfers the remaining population between the blocked polarized-triplet and $T_0$-like populations. Any $T_0$-like population that does not convert during the mixing dwell is therefore transferred back into the blocked manifold by the next burst, producing the pronounced odd--even dependence on $n_\mathrm{rep}$. The square shows the experimental initialization point, obtained with a randomized input using 11 repetitions and a \SI{1}{\micro\second} mixing dwell. Only about half of the initial population is blocked, whereas the calculated contours assume a fully polarized input and the minimum mixing dwell required to reach 99\% at each repetition number. Fixed pulse-transport overheads and the final verification readout are not included in $T_\mathrm{init}$.
}
\label{fig_ext:fig4}
\end{figure}

\end{document}